\documentclass[12pt]{article}
\pdfoutput=1
\usepackage{amsmath,amssymb}
\usepackage{graphicx}
\usepackage{color}
\usepackage{cases}
\usepackage[bookmarksnumbered=true,colorlinks=true,linkcolor=black,citecolor=black]{hyperref}
\newlength{\extraspace}
\newlength{\extraspaces}
\def\bsklength{.8mm} 

\newcommand{\beq}{\begin{equation}}
\newcommand{\eeq}{\end{equation}}
\newcommand{\bseq}{\addtocounter{subeqno}{1}\begin{subequations}}
\newcommand{\eseq}{\end{subequations}}

\newcommand{\newsection}[1]{
\vspace{6mm}
\pagebreak[3]
\addtocounter{section}{1}
\setcounter{subsection}{0}
\setcounter{figure}{0}
\phantomsection%
\addcontentsline{toc}{section}{\protect\numberline{\arabic{section}.}{#1}}
\noindent{\large \bf \thesection. #1}
\nopagebreak
\medskip
\nopagebreak}

\newcommand{\appsection}[1]{
\vspace{5mm}
\pagebreak[3]
\addtocounter{section}{1}
\setcounter{equation}{0}
\setcounter{subsection}{0}
\setcounter{figure}{0}
\phantomsection%
\addcontentsline{toc}{subsection}{\protect\numberline{\thesection.\ \ }{#1}}
\noindent{\large \bf \thesection. #1}
\nopagebreak
\medskip
\nopagebreak}

\font\mathscript=eusm10 at 12pt
\font\mathscripts=eusm7
\font\mathscriptss=eusm5
\newfam\mathscri
\textfont\mathscri=\mathscript
\scriptfont\mathscri=\mathscripts
\scriptscriptfont\mathscri=\mathscriptss
\def\mathscr#1{{\fam\mathscri\relax#1}}

\font\mathfrakt=eufm10 at 12pt
\font\mathfrakts=eufm7
\font\mathfraktss=eufm5
\newfam\mathfraki
\textfont\mathfraki=\mathfrakt
\scriptfont\mathfraki=\mathfrakts
\scriptscriptfont\mathfraki=\mathfraktss
\def\mathfrak#1{{\fam\mathfraki\relax#1}}

\def\CL{{\cal L}}

\def\T{{\sf T}}

\def\S{{\sf S}}

\renewcommand{\tilde}{\widetilde}

\def\Det{{\rm Det}}
\renewcommand{\bar}{\overline}

\def\half{{\tfrac{1}{2}}}

\newcommand{\gsim}{\gtrsim}

\begin{document}
\setcounter{page}{0}
\addtolength{\baselineskip}{\bsklength}
\thispagestyle{empty}
\renewcommand{\thefootnote}{\fnsymbol{footnote}}        

\begin{flushright}
arXiv:yymm.nnnn [hep-ph]\\
\end{flushright}
\vspace{.4cm}

\begin{center}
{\Large
{\bf{Prediction of Neutrino Mixing based on $C_2\times D_3$}}}\\[1.2cm]
{\rm HoSeong La\footnote{hsla.avt@gmail.com}
}
\\[3mm]
{\it Department of Physics and Astronomy,\\[1mm]
Vanderbilt University,\\[1mm]              
Nashville, TN 37235, USA} \\[1.5cm]

\vfill
{\parbox{15cm}{
\addtolength{\baselineskip}{\bsklength}
\noindent
The lepton mixing angles of the PMNS matrix are predicted 
based on the lepton flavor symmetry of a finite group $C_2\times D_3$, where
the cyclic group $C_2$ acts on the charged lepton mass terms and 
the dihedral group $D_3$ on the neutrino ones.
All three mixing angles of the PMNS matrix are given in terms of just one 
parameter, the charged lepton mixing angle, 
and fit extremely well to the observed values.
In particular, the smallness of $\theta_{13}$ is explained in terms of
the smallness of the muon-to-tau mass ratio. 

\bigskip
Keywords: neutrinos, PMNS matrix, discrete lepton flavor symmetry\\
PACS: 14.60.Pq, 11.30.Hv, 11.30.Er, 14.60.Lm   
}
}


\end{center}
\noindent
\vfill


\newpage
\setcounter{page}{1}
\setcounter{section}{0}
\setcounter{equation}{0}
\setcounter{footnote}{0}
\renewcommand{\thefootnote}{\arabic{footnote}}  
\newcounter{subeqno}
\setcounter{subeqno}{0}
\setlength{\parskip}{2mm}
\addtolength{\baselineskip}{\bsklength}

\pagenumbering{arabic}


\newsection{Introduction}

The success of the Standard Model (SM) of the Electroweak (EW) theory 
and Quantum Chromodynamics (QCD) is quite impressive as the last loophole
is finally closed with the long-waited discovery of the Higgs at LHC.
Nevertheless, it is equally well known that the SM is not our ultimate
theory to describe the Nature. 
The observations of neutrino flavor violations\cite{pdg}\cite{deGouvea:2013onf}
unambiguously supports the existence of physics beyond the SM. 
These observations are commonly interpreted to indicate that 
neutrinos are massive and the flavor-violations are due to the difference
between the flavor and mass eigenstates of the 
leptons\cite{Pontecorvo:1957qd}\cite{no_ref}.
For massive neutrinos, right-handed neutrinos are inevitably required and 
the mystery of the neutrino physics originates from the fact that the 
right-handed neutrinos do not carry any charges of the SM, 
but we have to rely on their visible companions of their left-handed partners
to investigate their properties. 
The only connection between the left-handed
and right-handed neutrinos are via their Dirac mass terms 
(i.e. Yukawa interactions involving the SM Higgs or something similar), 
unless there is a new physics in which the right-handed neutrinos 
also carry some new gauge
charges even below the EW scale\cite{La:2012ky}\cite{La:2013gga}.
So, to go beyond the SM it is important to understand
the structure of these neutrino masses and mixings. 
If there is any nontrivial structure which can be traced back to some
fundamental principle, it will be a clue to the physics beyond SM.
Partly motivated by this, in this paper we will present mass matrices 
based on new symmetry constraints, which explain the observed mixing 
angles.\footnote{Explaining the neutrino mixing based on a discrete symmetry 
was initiated in \cite{Harrison:2002er}.}

The relevant terms for the interactions between neutrinos 
and charged leptons in the SM Lagrangian are given by
\beq
\CL_{\ell\nu} 
= {g\over \sqrt{2}} W_\mu^+ \bar{\ell'_L}\gamma^\mu \nu'_{L}+{\rm h.c.}
= {g\over \sqrt{2}} W_\mu^+ 
\bar{\ell_L} U_\ell^\dagger\gamma^\mu U_\nu\nu_{L}+{\rm h.c.},
\eeq
where $\ell'$ and $\nu'$ denote the Weak flavor eigenstates of the
charged leptons
and neutrinos, respectively, $\ell$ and $\nu$ denote their mass eigenstates,
and $U_\ell$ is the theoretical (unitary) mixing matrix of the charged leptons
and $U_\nu$ is that of the left-handed neutrinos.
Since the existence of the neutrinos is only indirectly inferred by observing
the companion charged leptons, the observed neutrino
(to be precise, lepton) mixing is given by the celebrated PMNS 
matrix\cite{Pontecorvo:1957qd}\cite{no_ref}
\beq
\label{e:1}
U_{\rm PMNS}(\theta_{23},\theta_{13},\theta_{12})
=U_\ell^\dagger(\theta_{\ell 23},\theta_{\ell 13},\theta_{\ell 12})\,
U_\nu(\theta_{\nu 23},\theta_{\nu 13},\theta_{\nu 12}).
\eeq
Since three dimensional rotations satisfy the SO(3) group symmetry, 
no matter how many angles are involved on the r.h.s., 
only three Euler angles are needed to express the final rotation for 
the PMNS matrix, and these three angles are experimentally measured. 

The latest best-fit numbers from the pdgLive\cite{pdg2013} are
\footnote{No error range for $\sin^2(2\theta_{23})$ is posted yet, 
but it is not going to be really crucial for us since our predicted numbers 
are almost 0.95.}:
$\sin^2(2\theta_{12})=0.857{+0.023-0.025}$, $\sin^2(2\theta_{23})>0.95$,
and $\sin^2(2\theta_{13})=0.095\pm 0.010$, which can also be expressed as
\beq
\begin{aligned}
\label{e:s4bf}
\sin^2\theta_{12} &\simeq 0.311 \pm 0.016,\\ 
\sin^2\theta_{23}&\simeq (0.39\ {\sim}\ 0.61), \\
\sin^2\theta_{13}&\simeq 0.024 \pm 0.003.
\end{aligned}
\eeq
The masses are not completely determined, but two constraints are known:
\beq
\label{e:s4mbf}
\Delta m_{21}^2 \simeq 7.50{+0.19 \atop -0.20}\times 10^{-5}\ {\rm eV}^2,\quad
\Delta m_{32}^2 \simeq 2.32{+0.12 \atop -0.08}\times 10^{-3}\ {\rm eV}^2.
\eeq
In principle, there can be three different CP-violating phases for 
three generations of leptons, but in the Dirac case a $3\times 3$ unitary matrix allows only one phase to be independent and the rest can be eliminated
by chiral phase transformations.
However, in this paper we will ignore the CP-violation for simplicity of
the argument. The generalization should be straightforward, and we will 
comment on it in the discussion section later.

We emphasize that there is no {\it a priori} reason to demand $U_\ell=1$.
In fact, allowing more general $U_\ell$ in this paper, we can easily justify 
the measured PMNS matrix based on the new symmetry argument. 
We may even call it a prediction since no free parameter is involved 
once the right symmetry is imposed.
The new symmetry we introduce here is based on the cyclic groups $C_2$,
and enhanced to the dihedral group $D_3$\cite{hamgp}. 
The case of pure cyclic groups is presented in \cite{c23} and in this paper
we will show what happens if the symmetry is enhanced to the dihedral group 
and, at the same time, some technical details will be explained.
The symmetry we have introduced have not been 
previously considered, although many other discrete symmetries are 
investigated for the neutrino masses and mixing (see, for example,
a recent review \cite{King:2013eh} and references therein; also see
\cite{Altarelli:2010gt};
some mathematical details can be found in \cite{Grimus:2011fk}).

This paper is organized as follows. In section 2, we introduce a new
discrete symmetry based on finite groups. 
In section 3, the details of the $C_2$ symmetry for the charged lepton
mass matrix is explained, and the charged lepton mixing matrix $U_\ell$ 
is constructed. In section 4, the details of the $C_2\times C_2$ symmetry 
for the neutrinos are explained and the condition toward the enhanced $D_3$
symmetry is given.
Also the theoretical neutrino mixing matrix $U_\nu$ is constructed. 
Then in section 5, the PMNS matrix is given numerically and compared to 
the best-fit values.
Finally, the conclusions and discussions are given in the final section.

\newsection{The Symmetry}


The mixing matrices are dictated by the structure of the mass matrices.
So the symmetry we need is not only the symmetry of the Lagrangian 
but also what constrains the mass matrices to the desired form
at the same time. Thus, first, we can demand that
the mass matrices are invariant as
\begin{subequations}
\begin{align}
\label{e:7a}
{\T'}^\dagger \mathbf{M} \T' &= \mathbf{M}, \\
\label{e:7b}
\T^\dagger \mathbf{m}_{\rm D} \S  &= \mathbf{m}_{\rm D}, \\
\label{e:7c}
\S^\dagger \mathbf{m}_R \S^* &= \mathbf{m}_R,
\end{align}
\end{subequations}
where $\T'$, $\T$, and $\S$ are (nontrivial) operations 
acting on the charged leptons, left-handed neutrinos $\nu_L$, 
right-handed neutrinos $\nu_R$, respectively.
For $\T'$, $\T$, and $\S$ to be interpreted as symmetry transformations,
they must form a group. Since they can be all independent, the simplest group
we can consider is a direct product of three cyclic groups $C_2$ of order 
two\cite{hamgp}. $C_2$ is a group of just one nontrivial element in addition 
to the identity, a simplest possible nontrivial group. 
So, if ${\T'}^2=1$, ${\T}^2=1$, and $\S^2=1$, then 
we have a symmetry group $C_2(\T')\times C_2(\T)\times C_2(\S)$.
For them to be a symmetry of the Lagrangian,
$\T'$, $\T$, and $\S$ must be also unitary, and that they are also hermitian.
Then the discrete symmetry under
$C_2(\T')\times C_2(\T)\times C_2(\S)$ is a global 
(lepton flavor) symmetry of the Lagrangian.

In principle, we could demand different symmetry transformations 
for the left-handed and right-handed charged leptons, but as we will see later, 
even if we impose eq.(\ref{e:7a}) differently as 
${\T'}^\dagger \mathbf{M} \T'_R = \mathbf{M}$,
${\T'_R}^\dagger={\T'_R}$ for the desired $\mathbf{M}$, eq.(\ref{e:2a}), 
will dictate that ${\T'_R}={\T'}$ has to be satisfied.
One may also wonder if $\T'$ and $\T$ should be the same because left-handed
leptons form SU(2) doublets, but it is not necessary because their 
transformations are independent from the SU(2) isospin rotations, 
which are always compensated by those of the Higgs doublet. 
In some sense, this is also why neutrino mixing and 
charged lepton mixing can be different even though they form SU(2) doublets.
Since $\T'$, $\T$, and $\S$ have eigenvalues
$\pm 1$, they can be regarded as generalized parity operators analogous
to $\gamma_5$. $\T'$, $\T$, and $\S$ also fix 
mass matrices to the desired forms, as we will see later.

Furthermore, when $C_2(\T)\times C_2(\S)$ is enhanced to the dihedral group
$D_3(\T, \S)$, the entire mixing matrices will be fixed by the symmetry 
without an additional free parameter.
The dihedral group $D_3(\T, \S)$ can be presented by
\beq
\label{e:49}
D_3(\T, \S)=\langle\T, \S \,| \T^2=1, \S^2=1, (\T\S)^3=1\rangle.
\eeq
Geometrically, $D_3$ is also known as a symmetry of a regular triangle in the
two-dimensions, but we consider $D_3(\T, \S)$ in the three-dimensions.

For notational convenience, we first define three basic real symmetric operators
\beq
\label{e:g3}
\Gamma_1(\theta_1)
=\left(
\begin{array}{ccc}
1 & 0 &0\\ 
0 &c_1  &s_1\\
0 &s_1 &-c_1
\end{array}
\right), 
\quad
\Gamma_2(\theta_2)
=\left(
\begin{array}{ccc}
c_2 & 0 &s_2\\ 
0 &1  &0\\
s_2& 0 &-c_2
\end{array}
\right), 
\quad
\Gamma_3(\theta_3)
=\left(
\begin{array}{ccc}
c_3 &s_3& 0 \\ 
s_3 &-c_3 & 0\\
0  &0&1 
\end{array}
\right), 
\eeq
where we call $c_i\equiv \cos\theta_i$ or $s_i\equiv \sin\theta_i$ 
the symmetry parameters, and $\Gamma_i$ acts about $i$-axis. 
$\Gamma_i$'s are combinations of a
rotation and an inversion, satisfying $\Gamma_i^2=1$, and $\Det\Gamma_i=-1$.
Because of the latter condition, $\Gamma_i$'s are not group elements of SO(3)
or SU(3), hence the finite groups we consider here are not their subgroups.
In the followings, all $\T'$, $\T$, and $\S$ can be expressed in terms of these, 
so these are the building blocks of our cyclic groups and the dihedral group.


As was first observed in \cite{La:2013gga}, the current best-fit mixing angles 
can be extremely well reproduced if we parametrize 
$U_\ell(\theta_{\ell 13})$ and $U_\nu(\theta_{\nu 23}, \theta_{\nu 12})$
such that
\beq
\label{e:1a}
U_{\rm PMNS}(\theta_{23}, \theta_{13}, \theta_{12})
= U_\ell^\dagger(\theta_{\ell 13})\, U_\nu(\theta_{\nu 23}, \theta_{\nu 12})
\eeq
with respect to the mass matrices (in the Weak flavor basis) of the form
\bseq
\begin{align}
\label{e:2a}
\mathbf{M} &=
\left(
\begin{array}{ccc}
M_{11} & 0 &M_{13}\\ 
0 &M_{22}  &0\\
M_{13} &0 &M_{33}
\end{array}
\right),\\
\label{e:2b}
\mathbf{m}_{\rm D} &=
\left(
\begin{array}{ccc}
m_{11} & m_{12} &m_{13}\\ 
m_{12} &m_{22}  &0\\
m_{13} &0 &m_{22}
\end{array}
\right), 
\end{align}
\eseq
where $M_{22}=M_\mu$ is the muon mass for the charged lepton mass 
matrix $\mathbf{M}$ and $\mathbf{m}_{\rm D}$ is the Dirac neutrino mass matrix
with $m_{33}=m_{22}$. So, motivated by \cite{La:2013gga}, even without 
the extra ${\rm U(1)}_\lambda$ symmetry, we demand 
these mass matrices to be hermitian.
The specific form of the Majorana mass matrix for the right-handed neutrinos 
are not crucial in our argument, so we will only invoke later when we need it.
Now, our mission is to show the existence of proper $\T'$, $\T$, and $\S$, 
which constrain the mass matrices to be the desired forms given here.

\newsection{Charged Leptons}

Our plan is as follows: We will first construct the symmetry, then
show that the symmetry indeed constrains the mass matrix uniquely to
the desired form. Next, we will diagonalize the desired mass matrix 
to obtain the mixing matrix with the mixing angle constrained by the 
symmetry parameter.


The symmetry $C_2(\T')$ we need is generated explicitly by
\beq
\label{e:8a}
\T'=
\left(
\begin{array}{ccc}
{\tilde{M}_{33}-\tilde{M}_0\over \tilde{M}_{33}+\tilde{M}_0} & 0 
&{2M_{13}\over \tilde{M}_{33}+\tilde{M}_0}\\ 
0 &1 &0\\
{2M_{13}\over \tilde{M}_{33}+\tilde{M}_0} &0 
&-{\tilde{M}_{33}-\tilde{M}_0\over \tilde{M}_{33}+\tilde{M}_0}
\end{array}
\right),
\eeq
where we have introduced a shorthand notation
\beq
\label{e:8ax}
\tilde{M}\equiv M-M_e,
\eeq
such that the required cyclic condition $\T'^2=1$ is satisfied provided
\begin{equation}
\label{e:11}
M_{13}^2=\tilde{M}_0 \tilde{M}_{33}.
\end{equation}
In terms of the notations introduced in eq.(\ref{e:g3})
this $\T'$ can be expressed as
\begin{equation}
\label{e:8}
\T'=\Gamma_2(\theta_\ell),
\end{equation}
where the symmetry parameters are given by
\bseq
\begin{align}
\label{e:8p}
c_\ell &={\tilde{M}_{33}-\tilde{M}_0\over \tilde{M}_{33}+\tilde{M}_0},\\
\label{e:8q}
s_\ell &=-{2M_{13}\over \tilde{M}_{33}+\tilde{M}_0}
\end{align}
\eseq
and $c_\ell^2+s_\ell^2=1$ because of eq.(\ref{e:11}).

\noindent
\underline{Uniqueness of $\mathbf{M}$}

We can now show that eq.(\ref{e:7a}) indeed leads $\mathbf{M}$
to be the desired form of eq.(\ref{e:2a}) with $M_{11}=M_0$. 
For this purpose, we will first consider a more 
general $\mathbf{M}=(M_{\ell\ell'})$, 
then eq.(\ref{e:7a}) reads
\beq
\label{e:8r}
\left(
\begin{array}{ccc}
c_\ell^2 M_{11}+2c_\ell s_\ell M_{13}+s_\ell^2 M_{33} 
& c_\ell M_{12}+s_\ell M_{23} 
&c_\ell s_\ell (M_{11}-M_{33})+(s_\ell^2-c_\ell^2)M_{13}\\ 
c_\ell M_{12}+s_\ell M_{23} &M_{22} & s_\ell M_{12}-c_\ell M_{23}\\
c_\ell s_\ell (M_{11}-M_{33})+(s_\ell^2-c_\ell^2)M_{13} 
&s_\ell M_{12}-c_\ell M_{23} 
&s_\ell^2 M_{11}-2c_\ell s_\ell M_{13}+c_\ell^2 M_{33}
\end{array}
\right)
=(M_{\ell\ell'}),
\eeq
which consists of six linear equations for six mass parameters 
$M_{\ell\ell'}=M_{\ell' \ell}$.
The (13)-components lead to
\beq
\label{e:8s}
t_\ell\equiv \tan\theta_\ell=-{2M_{13}\over M_{33}-M_{11}},
\eeq
which (11)- and (33)-components also satisfy.
Comparing this to eqs.(\ref{e:8p})(\ref{e:8q}), $M_{11}$ can be fixed as
\beq
\label{e:8sx}
M_{11}=M_0.
\eeq
(12)-components (also (23)-components) lead to
\beq
\label{e:8t}
M_{23}={1-c_\ell\over s_\ell}M_{12}=-{\tilde{M}_0\over M_{13}} M_{12}.
\eeq
Since the three eigenvalues of $\mathbf{M}$ are the physical charged lepton
masses $M_e$, $M_\mu$, and $M_\tau$, 
\beq
\label{e:8u}
\Det(\mathbf{M}-\lambda\mathbf{1})
=(M_e-\lambda)(M_\mu-\lambda)(M_\tau-\lambda).
\eeq
With eq.(\ref{e:8sx}), for $\lambda=M_e$ this becomes
\beq
\label{e:8v}
0=\tilde{M}_{22}\left(\tilde{M}_0\tilde{M}_{33}-M_{13}^2\right)
-M_{12}^2{(\tilde{M}_{33}+\tilde{M}_0)^2\over \tilde{M}_{33}}.
\eeq
The first term vanishes because of eq.(\ref{e:11}), then the remaining
term implies
\beq
\label{e:8w}
M_{12}=0
\eeq
and that $M_{23}=0$ can be shown from eq.(\ref{e:8t}). 
Similarly, for $\lambda=M_\mu$,
we can easily obtain $M_{22}=M_\mu$, then $\lambda=M_\tau$ is also satisfied
with eqs.(\ref{e:11})(\ref{e:8w}).
Thus, the symmetry $C_2(\T')$ uniquely fixes the charged lepton mass
matrix to the desired form given in eq.(\ref{e:2a}) with $M_{11}=M_0$.

\noindent
\underline{Proof of $\T'_R=\T'$}

Next, we will show $\T'_R=\T'$ for
${\T'}^\dagger \mathbf{M} \T'_R = \mathbf{M}$,
if $\mathbf{M}$ is given by eq.(\ref{e:2a}). Let us start with
\beq
\label{e:9x}
\T'_R=\left(
\begin{array}{ccc}
T'_{11} & 0 &T'_{13}\\ 
0 &1  &0\\
T'_{31} &0 &-T'_{11}
\end{array}
\right)
\eeq
which is not necessarily symmetric and satisfies
\begin{equation}
\label{e:9}
\T'_R= \mathbf{M}^{-1} \T' \mathbf{M}.
\end{equation}
Then we can easily see ${\T'}_R^2=1$, Since $\T'^2=1$. 

Next we need to make sure $\T'_R$ is also hermitian or real-symmetric.
The components of $\T'_R$ can be read off explicitly from the above as
\bseq
\begin{align}
\label{e:M12a}
T'_{11} & ={M_\mu\over\Det \mathbf{M}}\left(
c_\ell(M_{11}M_{33}+M_{13}^2) +s_\ell M_{13}(M_{33}-M_{11})\right), \\
\label{e:M12b}
T'_{13} & ={M_\mu\over\Det \mathbf{M}}\left(
2c_\ell M_{13}M_{33} +s_\ell (M_{33}^2-M_{13}^2)\right), \\
\label{e:M12c}
T'_{31} & ={M_\mu\over\Det \mathbf{M}}\left(
-2c_\ell M_{11}M_{13} +s_\ell (M_{11}^2-M_{13}^2)\right).
\end{align}
\eseq
To obtain symmetric $\T'_R$, we demand that $T'_{13}=T'_{31}$, 
which leads to eq.(\ref{e:8s}), and that $T'_{11}=c_\ell$ and $T'_{13}=s_\ell$.
Thus we obtain $\T'_R=\T'$. So the left- and right-handed charged
leptons transform in the same way and
eq.(\ref{e:7a}) is good enough for our purpose.

\noindent
\underline{Obtaining $U_\ell$}

Having established that the charged lepton mass matrix $\mathbf{M}$ given in 
eq.(\ref{e:2a}) indeed is what we get under the symmetry $C_2(\T')$, 
we can now diagonalize it as
\beq
U_\ell^{-1} \mathbf{M}\, U_\ell=\mathbf{M}_{\rm d}
={\rm diag}(M_e, M_\mu, M_\tau)
\eeq
to obtain the charged lepton mixing matrix
\beq
\label{e:10}
U_\ell(\theta_{\ell 13}) = 
\left(
\begin{array}{ccc}
c_{\ell 13} & 0 &-s_{\ell 13}\\ 
0 &1 &0\\
s_{\ell 13} &0 &c_{\ell 13}
\end{array}
\right),
\eeq
where the mixing angle satisfies
\bseq
\begin{align}
\label{e:10xa}
s_{\ell 13} &=-{M_{13}\over \sqrt{(M_{33}-M_e)^2 +M_{13}^2}},\\
\label{e:10xb}
c_{\ell 13} &={M_{33}-M_e\over \sqrt{(M_{33}-M_e)^2 +M_{13}^2}}.
\end{align}
\eseq
In addition to $M_\mu=M_{22}$, the other eigenvalues are now given by
\bseq
\begin{align}
\label{e:26x}
M_e &=\half\left(M_{33}+M_{11}-\sqrt{(M_{33}-M_{11})^2+4M_{13}^2}\right), \\
\label{e:26y}
M_\tau &=\half\left(M_{33}+M_{11}+\sqrt{(M_{33}-M_{11})^2+4M_{13}^2}\right).
\end{align}
\eseq
Note that eq.(\ref{e:26x}) is actually equivalent to eq.(\ref{e:11}).
Since $M_e$ and $M_\tau$ are given by three parameters $M_{11}$, $M_{33}$ and
$M_{13}$, one of which is a free parameter. This free parameter is 
constrained by $c_\ell$ so that fixing $c_\ell$ in terms of $M_0$ can 
fix $M_{11}$, and vice versa. 

Together with $M_\tau$, eq.(\ref{e:10xa}) can be expressed as
\beq
\label{e:10x}
s_{\ell 13}=\sqrt{M_{11}-M_e\over M_\tau -M_e}.
\eeq
This is interesting because it is almost given in terms of physical charged
lepton masses except $M_{11}$. So, it certainly motivates us to choose 
$M_{11}=M_\mu$ so that it can be entirely given in terms of all three
physical charged lepton masses. This can be achieved by choosing the symmetry
of the generator $\T'$ given in terms of $M_0=M_\mu$, which will fix 
$M_{11}=M_\mu$. Therefore, a good $C_2(\T)$ symmetry motivated by phenomenology 
is to let $M_0=M_\mu$, as in \cite{La:2013gga},
so that $s_{\ell 13}$ is given by the muon-to-tau mass ratio to a good
approximation. As we will see later, this leads to neutrino mixing angles
very close to the best-fit values.

There is also an interesting relationship between the symmetry parameter 
$\theta_\ell$ and the mixing angle $\theta_{\ell 13}$.
Comparing eq.(\ref{e:10xa}) to eq.(\ref{e:8p}), we can show that
\beq
\label{e:10y}
1-2 s_{\ell 13}^2= {\tilde{M}_{33}^2-M_{13}^2\over \tilde{M}_{33}^2 +M_{13}^2}
=c_\ell,
\eeq
where eq.(\ref{e:11}) is used for the last equality. This leads to an 
interesting identity
\beq
\label{e:10z}
\theta_\ell=2\theta_{\ell 13}.
\eeq
Thus the symmetry angle of $C(\T')$ is actually twice of the mixing angle
so that the charged lepton mixing angle $\theta_{\ell 13}$ can play 
an important role in our model.

\newsection{Neutrinos}

For neutrinos, we will proceed similarly as the charged lepton case but
with some alterations.
First, we construct the simpler symmetry based on the cyclic groups, 
and show the uniqueness of the form of the mass matrix. However, since
$m_{11}=m_{22}$ is a special case, we will treat $m_{11}\neq m_{22}$
and $m_{11}=m_{22}$ cases separately.
Next, we will look for a condition to enhance the symmetry group to 
the dihedral group, in which the symmetry parameters of the cyclic groups
will be further constrained. Then we will construct the mixing matrix,
in which now mixing angles will be constrained by the symmetry parameters.

For the symmetry $C_2(\T)$ of the left-handed neutrinos, 
we choose, using the notation given in 
eq.(\ref{e:g3}),
\begin{equation}
\label{e:8x}
\T=\Gamma_2(\theta_\nu),
\end{equation}
where $\theta_\nu$ should be properly chosen to satisfy our need.
In principle, we can choose any value for $c_\nu$ because $\T^2=1$ for 
any $c_\nu$. However, if we want to relate it to the values in the 
charged lepton case, an interesting choice is
\begin{equation}
\label{e:8y}
c_\nu=t_{\ell 13}=-{M_{11} -M_e\over M_{13}}.
\end{equation}
Although this is not necessarily a unique choice, but certainly a good choice 
so that $\theta_\nu$ can be related to $\theta_{\ell 13}=\theta_\ell/2$. 
It will be extremely interesting if there exists a symmetry argument to
force this identity, which we will leave as future work.

Once $\T$ is chosen, $\S$ for the right-handed neutrinos should be what can 
constrain the Dirac neutrino mass matrix to be the desired form given 
in eq.(\ref{e:2b}). 
To find such $\S$, let us start with
\beq
\label{e:34u}
\S =\Gamma_1(\theta_\alpha) \Gamma_3(\theta_\nu) \Gamma_1(\theta_\alpha)
=\left(
\begin{array}{ccc}
c_\nu & s_\nu c_\alpha &s_\nu s_\alpha\\ 
s_\nu c_\alpha &s_\alpha^2-c_\nu c_\alpha^2& -c_\alpha s_\alpha (1+c_\nu)\\
s_\nu s_\alpha &-c_\alpha s_\alpha (1+c_\nu) &c_\alpha^2-c_\nu s_\alpha^2
\end{array}
\right).
\eeq
Note that $\S^2=1$ and $\S^\dagger=\S$ which we can easily see from
eq.(\ref{e:g3}) so that $\S$ certainly forms the cyclic group 
$C_2(\S)$ which satisfies our criteria to be a symmetry of the Lagrangian.
Then, $\theta_\alpha$ can be determined specifically 
to satisfy eq.(\ref{e:7b}), which we will show next.

\noindent
\underline{$m_{11}\neq m_{22}$ Case}

Let us first consider a more general Dirac neutrino mass matrix, 
eq.(\ref{e:2b}), with $m_{11}\neq m_{22}$.
With our candidate representation of $\S$ given in eq.(\ref{e:34u}), 
the symmetry invariance constraint eq.(\ref{e:7b}) in components reads 
\bseq
\begin{align}
\label{e:38a}
m_{11} &= c_\nu (c_\nu m_{11}+s_\nu m_{13})+c_\nu s_\nu c_\alpha m_{12}
+ s_\nu s_\alpha (c_\nu m_{13}+s_\nu m_{22}), \\
\label{e:38b}
m_{12} &= s_\nu c_\alpha (c_\nu m_{11}+s_\nu m_{13})
+c_\nu (s_\alpha^2 -c_\nu c_\alpha^2) m_{12}
- c_\alpha s_\alpha (1+c_\nu)(c_\nu m_{13}+s_\nu m_{22}), \\
\label{e:38c}
m_{12} &= c_\nu m_{12}+s_\nu c_\alpha m_{22},\\
\label{e:38d}
m_{22} &=s_\nu c_\alpha m_{12} +(s_\alpha^2 -c_\nu c_\alpha^2) m_{22}, \\
\label{e:38e}
m_{13} &= s_\nu s_\alpha (c_\nu m_{11}+s_\nu m_{13})
- c_\alpha s_\alpha (1+c_\nu) c_\nu m_{12}
+(c_\alpha^2 -c_\nu s_\alpha^2) (c_\nu m_{13}+s_\nu m_{22}), \\
\label{e:38f}
m_{13} &=c_\nu (s_\nu m_{11}-c_\nu m_{13}) +s_\nu^2 c_\alpha m_{12}
+s_\nu s_\alpha (s_\nu m_{13}-c_\nu m_{22}), \\
\label{e:38g}
0=m_{23} &=s_\nu s_\alpha m_{12} -c_\alpha s_\alpha (1+c_\nu) m_{22}, \\
\label{e:38h}
0=m_{23} &= s_\nu c_\alpha (s_\nu m_{11}-c_\nu m_{13})
+s_\nu (s_\alpha^2 -c_\nu c_\alpha^2) m_{12}
- c_\alpha s_\alpha (1+c_\nu)(s_\nu m_{13}-c_\nu m_{22}), \\
\label{e:38i}
m_{22} &= s_\nu s_\alpha (s_\nu m_{11}-c_\nu m_{13})
- c_\alpha s_\alpha (1+c_\nu) s_\nu m_{12}
+(c_\alpha^2 -c_\nu s_\alpha^2) (s_\nu m_{13}-c_\nu m_{22}).
\end{align}
\eseq
Eqs.(\ref{e:38c})(\ref{e:38d})(\ref{e:38g}) lead to 
(for $c_\alpha\neq 0 \neq s_\alpha$)
\beq
\label{e:39}
{m_{12}\over m_{22}} ={s_\nu c_\alpha\over 1-c_\nu}
={(1+c_\nu) c_\alpha\over s_\nu}.
\eeq
Eq.(\ref{e:38a})$\times (c_\nu/s_\nu) +$eq.(\ref{e:38f}) leads to
\beq
\label{e:40}
{m_{13}\over m_{12}} ={c_\alpha\over 1-s_\alpha}.
\eeq
With eqs.(\ref{e:39})(\ref{e:40}),
eq.(\ref{e:38a})$\times s_\nu - c_\nu\times$eq.(\ref{e:38f}) leads to
\beq
\label{e:41}
{m_{12}\over m_{11}} ={s_\nu c_\alpha \over 2c_\nu +s_\alpha(1+c_\nu)}.
\eeq
The rest of equations equivalently reproduce these, so 
eqs.(\ref{e:39})-(\ref{e:41}) are complete constraints 
obtained from the symmetry constraint eq.(\ref{e:7b}) 
for $C_2(\T)\times C_2(\S)$. So the mass parameters are constrained by the 
symmetry parameters $\theta_\nu$ and $\theta_\alpha$.

The above relations between the mass parameters and the symmetry parameters
can be rearranged to show more useful relations.
From eqs.(\ref{e:39})(\ref{e:41}), eliminating $m_{12}$, we can obtain
a relation between $m_{11}$ and $m_{22}$ as
\beq
\label{e:42}
m_{11}-m_{22}
=\left({1+c_\nu\over 1-c_\nu}s_\alpha-{1-3c_\nu\over 1-c_\nu}\right) m_{22}.
\eeq
This in particular shows why $m_{11}=m_{22}$ is a special case,
for which $s_\alpha$ and $c_\nu$ become no longer independent.
If $m_{11}\neq m_{22}$, $s_\alpha$ and $c_\nu$ are independent.

Eq.(\ref{e:40}) also enables us to express $\theta_{\nu 23}$, 
one of the two neutrino mixing angles, in terms of a symmetry parameter as
\beq
\label{e:43a}
s_{\nu 23}^2 ={m_{13}^2\over m_{12}^2+m_{13}^2}={1+s_\alpha \over 2},
\eeq
where the former identity is from eq.(\ref{e:26bb}).
Then together with eqs.(\ref{e:39})(\ref{e:42}), we can obtain the other
neutrino mixing angle $\theta_{\nu 12}$,
depending on the choice of $m_1$, as
\bseq
\begin{numcases}
{{1-c_\nu\over 1+c_\nu}{2\over 1+s_\alpha}
={m_{12}^2+m_{13}^2\over (m_+-m_{22})^2}
=}
t_{\nu 12}^2, \ \mbox{for } m_1=m_-, \label{e:43ba}\\[6pt]
{1\over t_{\nu 12}},\ \mbox{for } m_1=m_+, \label{e:43bb}
\end{numcases}
\eseq
where the latter identity is from eqs.(\ref{e:26bc})(\ref{e:26bd}) 
for $m_1=m_-$ or  eqs.(\ref{e:26bcx})(\ref{e:26bdx}) for $m_1=m_+$ and
\beq
\label{e:44}
{m}_{\pm}\equiv \half\left({m}_{11}+{m}_{22}
\pm\sqrt{({m}_{11}-{m}_{22})^2
+4({m}_{12}^2+{m}_{13}^2)}\right).
\eeq
So both mixing angles are now expressed in terms of the symmetry parameters,
$c_\nu$ and $s_\alpha$.
If we eliminate $s_\alpha$ from eqs.(\ref{e:43ba})(\ref{e:43bb})
with the help of eq.(\ref{e:43a}),
we can obtain interesting relationships between two mixing angles
in terms of $c_\nu$ as
\bseq
\begin{numcases}
{{1-c_\nu\over 1+c_\nu}=}
s_{\nu 23}^2 t_{\nu 12}^2\, ,\ \ \mbox{for } m_1=m_-, \label{e:45a}\\[6pt]
{s_{\nu 23}^2 \over t_{\nu 12}^2}\, ,\qquad
 \mbox{for } m_1=m_+. \label{e:45b}
\end{numcases}
\eseq

\noindent
\underline{$m_{11}= m_{22}$ Case}

As we have noticed in eq.(\ref{e:42}), $m_{11}=m_{22}$ can simplify
the symmetry parameters by relating each other.
This is also the case in which the Dirac neutrino mass matrix has identical 
diagonal components, i.e. diagonal components have maximal symmetry\cite{c23}.

From eq.(\ref{e:42}), if $m_{11}= m_{22}$, we can eliminate one symmetry
parameter using the other as
\beq
\label{e:34vb}
s_\alpha = {1-3c_\nu\over 1+c_\nu}, 
\eeq
then eq.(\ref{e:41}) simplifies as
\beq
\label{e:34va}
c_\alpha = {\sqrt{8c_\nu(1-c_\nu)}\over 1+c_\nu}
={m_{12}\over m_{11}}s_{\nu 23}.
\eeq
In particular, eq.(\ref{e:43a}) leads to
\beq
\label{e:45x}
{1-c_\nu\over 1+c_\nu}=s_{\nu 23}^2,
\eeq
hence eqs.(\ref{e:43ba})(\ref{e:43bb}) simplify to 
\beq
t_{\nu 12}^2=1.
\eeq
such that both eqs.(\ref{e:45a})(\ref{e:45b}) now reduce to eq.(\ref{e:45x}).
So, if we choose the symmetry parameter $s_\alpha$ as eq.(\ref{e:34vb}) 
in the beginning, we can force $m_{11}=m_{22}$ so that the symmetry 
can lead to $\theta_{\nu 12}=\pi/4$.

\noindent
\underline{Uniqueness of $m_{\rm D}$}

Once we have figured out how the symmetry angles 
$\theta_\alpha$ and $\theta_\nu$ 
satisfy the symmetry constraint eq.(\ref{e:7b}), it is straightforward 
to show how the symmetry eq.(\ref{e:7b}) constrains the Dirac neutrino mass
matrix $m_{\rm D}$ to be the desired form given in eq.(\ref{e:2b}). 
For this purpose, we only need to show that there are 
$s_\alpha$ and $c_\nu$ which set $m_{23}=0$ and $m_{33}=m_{22}$.

As in the charged lepton case, let us first start with a general hermitian
$m_{\rm D}=(m_{\ell\ell'})$, then the symmetry invariance condition
eq.(\ref{e:7b}) becomes
\beq
\begin{aligned}
\label{e:34v}
m_{\rm D}&=\T m_{\rm D} \S \\
&=\left(
\begin{array}{ccc}
c_\nu m_{11}+s_\nu m_{13}
& c_\nu m_{12}+s_\nu m_{23}
&c_\nu m_{13}+s_\nu m_{33}\\ 
m_{12} &m_{22}& m_{23}\\
s_\nu m_{11}-c_\nu m_{13}
& s_\nu m_{12}-c_\nu m_{23}
&s_\nu m_{13}-c_\nu m_{33}
\end{array}
\right)\\
&\ \ \times\left(
\begin{array}{ccc}
c_\nu & s_\nu c_\alpha &s_\nu s_\alpha\\ 
s_\nu c_\alpha &s_\alpha^2-c_\nu c_\alpha^2& -c_\alpha s_\alpha (1+c_\nu)\\
s_\nu s_\alpha &-c_\alpha s_\alpha (1+c_\nu) &c_\alpha^2-c_\nu s_\alpha^2
\end{array}
\right).
\end{aligned}
\eeq
The (21)-components read
\beq
\label{e:34w}
m_{12}=c_\nu m_{12}+s_\nu c_\alpha m_{22}+ s_\nu s_\alpha m_{23}.
\eeq
Now we choose the symmetry parameter for $\S$ to satisfy (see eq.(\ref{e:39}))
\beq
\label{e:34x}
c_\alpha =\left({1-c_\nu\over s_\nu}\right){m_{12}\over m_{22}},
\eeq
then (21)-components satisfy
\beq
\label{e:34y}
m_{23}
={1\over s_\alpha}\left({1-c_\nu\over s_\nu}m_{12}-c_\alpha m_{22}\right)=0.
\eeq

In fact, eq.(\ref{e:34x}) is the only condition we need to constrain 
even $m_{33}=m_{22}$.
The (11)- and (31)-components lead to eq.(\ref{e:40}) after eliminating
$m_{33}$ terms, then together with eq.(\ref{e:34x}), we can obtain
\beq
\label{e:34z}
{m_{13}\over m_{22}}={(1+c_\nu)c_\alpha^2\over s_\nu (1-s_\alpha)}.
\eeq
On the other hand, the (13)- and (33)-components independently lead to
\beq
\label{e:34za}
{m_{13}\over m_{33}}={(1+c_\nu)c_\alpha^2\over s_\nu (1-s_\alpha)}.
\eeq
Comparing these two, we can obtain
\beq
m_{33}=m_{22}.
\eeq
So, with the choice of eq.(\ref{e:34x}), the symmetry based on
$C_2(\T)\times C_2(\S)$ indeed constrains the Dirac neutrino mass matrix
$m_{\rm D}$ to be the desired form uniquely.

\noindent
\underline{$D_3$ Symmetry for $m_{11}\neq m_{22}$}

Under $C_2(\T)\times C_2(\S)$, $\theta_\alpha$ and $\theta_\nu$ are not
directly related but in terms of some mass parameters. (For example, see
eq.(\ref{e:39}).) So inevitably it will raise the question if the symmetry 
can be enlarged so that they can be more directly related.
This is indeed the case under certain conditions and the larger symmetry, 
as we will show in the following.

Using the definitions of $\T$ and $\S$ given in eqs.(\ref{e:8x})(\ref{e:34u}), 
we can compute
\beq
\label{e:46}
\T\S=
\left(
\begin{array}{ccc}
c_\nu^2+s_\alpha s_\nu^2 & 
s_\nu c_\alpha \left\{c_\nu- s_\alpha(1+c_\nu)\right\} &
s_\nu (c_\nu s_\alpha +c_\alpha^2-c_\nu s_\alpha^2)\\ 
s_\nu c_\alpha &s_\alpha^2-c_\nu c_\alpha^2  &-c_\alpha s_\alpha(1+c_\nu)\\
c_\nu s_\nu(1-s_\alpha) &
s_\nu^2 c_\alpha +  c_\alpha s_\alpha c_\nu(1+c_\nu)&
s_\nu^2 s_\alpha -c_\nu(c_\alpha^2-c_\nu s_\alpha^2)
\end{array}
\right).
\eeq
Now we need to check if
\beq
\label{e:47}
(\T\S)^3=1,\quad i.e.\ (\T\S)^2=\S\T.
\eeq
From the latter equation, with $\S\T=(\T\S)^\dagger$, the $(11)$-component reads
\begin{align}
0 &=(c_\nu^2+s_\alpha s_\nu^2)^2 
+s_\nu^2 c_\alpha^2 \left\{c_\nu- s_\alpha(1+c_\nu)\right\}
+c_\nu s_\nu^2 (1-s_\alpha)(c_\nu s_\alpha +c_\alpha^2-c_\nu s_\alpha^2)
-(c_\nu^2+s_\alpha s_\nu^2) \nonumber\\
\label{e:48}
&=s_\nu^2 (s_\alpha-1)\left\{(c_\nu+1)s_\alpha -c_\nu\right\}
\left\{(c_\nu+1)s_\alpha +2-c_\nu\right\}.
\end{align}
Note that $s_\alpha=\frac{c_\nu -2}{1+c_\nu}$ leads to undesirable 
$s_{\nu 23}^2$ in eq.(\ref{e:43a}), so not interesting to our purpose.
So, for $s_\nu\neq 0$ and $s_\alpha\neq 1$, as our solution, we take
\beq
\label{e:48x}
s_\alpha = \frac{c_\nu}{1+c_\nu} .
\eeq
Then, the following useful identities can be worked out:
\bseq
\begin{align}
\label{e:46a}
c_\nu^2+s_\alpha s_\nu^2 &=c_\nu^2 +{c_\nu s_\nu^2\over 1+c_\nu}
=c_\nu,\\
\label{e:46b}
c_\alpha^2-c_\nu s_\alpha^2 &=1-{c_\nu^2\over 1+c_\nu},\\
\label{e:46c}
s_\alpha^2-c_\nu c_\alpha^2 &=-s_\alpha.
\end{align}
\eseq
With these identities, eq.(\ref{e:46}) can be simplified as
\beq
\label{e:46x}
\T\S=
\left(
\begin{array}{ccc}
c_\nu & 0 & s_\nu \\ 
s_\nu c_\alpha &-s_\alpha &-c_\nu c_\alpha \\
s_\nu s_\alpha & c_\alpha& -c_\nu s_\alpha 
\end{array}
\right).
\eeq
Now we can easily show that
\beq
\label{e:46y}
(\T\S)^2=
\left(
\begin{array}{ccc}
c_\nu & s_\nu c_\alpha & s_\nu s_\alpha\\ 
0 &-s_\alpha & c_\alpha \\
s_\nu  & -c_\nu c_\alpha& -c_\nu s_\alpha 
\end{array}
\right)
=(\T\S)^\dagger=\S\T, \ \ i.e.\ (\T\S)^3=1.
\eeq
Therefore, with eq.(\ref{e:48x}), the symmetry $C_2(\T)\times C_2(\S)$ now 
gets enhanced to the dihedral group $D_3(\T, \S)$ with a presentation given 
by eq.(\ref{e:49}), under which $s_\alpha$ is fixed in terms of $c_\nu$. 
So, as the symmetry gets enhanced, we can eliminate another independent 
symmetry parameter, leaving the entire symmetry of the neutrino mass matrix 
controlled solely by $\theta_\nu$ now.

The mixing angles can now be expressed only in terms of $c_\nu$. 
Substituting $s_\alpha$ using eq.(\ref{e:48x}), eq.(\ref{e:43a}) 
becomes
\beq
\label{e:50a}
s_{\nu 23}^2 = {2c_\nu + 1\over 2(1+c_\nu)},
\eeq
and, from eqs.(\ref{e:43ba})(\ref{e:43bb}), we obtain
\bseq
\begin{numcases}
{t_{\nu 12}^2 =}{2(1-c_\nu)\over 2c_\nu + 1}, 
\quad \mbox{for } m_1=m_-, \label{e:50b}\\[6pt]
{2c_\nu+1\over 2(1-c_\nu)},\quad \mbox{for } m_1=m_+. \label{e:50c}
\end{numcases}
\eseq
So the neutrino mixing angles are entirely given in terms of the symmetry
parameter $\theta_\nu$ now.
Furthermore, eq.(\ref{e:42}) can be also given in terms of $c_\nu$ as
\beq
\label{e:51}
m_{11}-m_{22}={4c_\nu -1\over 1-c_\nu}m_{22}.
\eeq

\noindent
\underline{$D_3$ Symmetry for $c_\nu={1\over 4}$ and $m_{11}=m_{22}$}

From eq.(\ref{e:51}), if $m_{11}=m_{22}$, we get a simple fraction $c_\nu=1/4$. 
In fact, if we set eq.(\ref{e:50b}) and eq.(\ref{e:50c}) to be the same,
we also obtain $c_\nu=1/4$ and that $t_{\nu 12}^2=1$. So, the value
$c_\nu=1/4$ as well as $m_{11}=m_{22}$ are, in this sense, special,
calling for extra attention. 

We can also explicitly show in numbers that the dihedral symmetry condition,
$(\T\S)^3=1$, is satisfied as follows. The group generators are given
explicitly as
\bseq
\begin{align}
\T &=
\left(
\begin{array}{ccc}
\frac{1}{4} & 0 &{\sqrt{15}\over 4}\\ 
0 &1  &0\\
{\sqrt{15}\over 4} &0 &-\frac{1}{4}
\end{array}
\right), \\
\S &=
\left(
\begin{array}{ccc}
1 & 0 &0\\ 
0 &{\sqrt{24}\over 5}  &{1\over 5}\\
0 &{1\over 5} &-{\sqrt{24}\over 5}
\end{array}
\right)
\left(
\begin{array}{ccc}
{1\over 4} &{\sqrt{15}\over 4} &0\\ 
{\sqrt{15}\over 4} &-{1\over 4}  &0\\
0 &0 &1
\end{array}
\right)
\left(
\begin{array}{ccc}
1 & 0 &0\\ 
0 &{\sqrt{24}\over 5}  &{1\over 5}\\
0 &{1\over 5} &-{\sqrt{24}\over 5}
\end{array}
\right)
=\left(
\begin{array}{ccc}
{1\over 4} &{3\sqrt{10}\over 10} &{\sqrt{15}\over 20}\\ 
{3\sqrt{10}\over 10} &-{1\over 5}  &-{\sqrt{6}\over 10}\\
{\sqrt{15}\over 20} &-{\sqrt{6}\over 10} &{19\over 20}
\end{array}
\right), 
\end{align}
\eseq
then
\bseq
\begin{align}
\T\S &
=\left(
\begin{array}{ccc}
{1\over 4} &0 &{\sqrt{15}\over 4}\\ 
{3\sqrt{10}\over 10} &-{1\over 5}  &-{\sqrt{6}\over 10}\\
{\sqrt{15}\over 20} &{2\sqrt{6}\over 5} &-{1\over 20}
\end{array}
\right), \\
(\T\S)^2 &
=\left(
\begin{array}{ccc}
{1\over 4} &{3\sqrt{10}\over 10} &{\sqrt{15}\over 20}\\ 
0 &-{1\over 5}  &{2\sqrt{6}\over 5}\\
{\sqrt{15}\over 4} &-{\sqrt{6}\over 10} &-{1\over 20}
\end{array}
\right)
=(\T\S)^T=\S\T.
\end{align}
\eseq
So, under this $D_3(\T, \S)$, since five (out of total six)
mass parameters of $\mathbf{m}_{\rm D}$ except the overall mass scale
given by $m_{22}$ are fixed by the symmetry,
the neutrino mixing angles are to be completely fixed by the symmetry
without any additional free parameter left over.

\noindent
\underline{Diagonalizing $\mathbf{m}_{\rm D}$}

Having established the fact that the symmetry indeed constrains the 
Dirac neutrino mass matrix $\mathbf{m}_{\rm D}$ to be the desired form 
given in eq.(\ref{e:2b}), we can now diagonalize $\mathbf{m}_{\rm D}$ as
\beq
U_\nu^{-1} \mathbf{m}_{\rm D}\, U_\nu=\mathbf{m}_{\rm Dd}
={\rm diag}(m_1, m_2, m_3)
\eeq
with the theoretical neutrino mixing matrix
\beq
\label{e:26b}
U_\nu(\theta_{\nu 23}, \theta_{\nu 12}) =
\left(
\begin{array}{ccc}
c_{\nu 12} & s_{\nu 12} &0\\ 
-c_{\nu 23}s_{\nu 12} &c_{\nu 23}c_{\nu 12} &s_{\nu 23}\\
s_{\nu 23}s_{\nu 12} &-s_{\nu 23}c_{\nu 12} &c_{\nu 23}
\end{array}
\right),
\eeq
where $\theta_{\nu 23}$ and $\theta_{\nu 12}$ depend on the choice of 
$m_1$ and $m_2$ from $m_\pm$ given by
\beq
\label{e:26be}
{m}_{\pm}\equiv \half\left({m}_{11}+{m}_{22}
\pm\sqrt{({m}_{11}-{m}_{22})^2
+4({m}_{12}^2+{m}_{13}^2)}\right).
\eeq
If $m_1=m_-$, $\theta_{\nu 12}$ and $\theta_{\nu 23}$ are given by
\bseq
\begin{align}
\label{e:26ba}
c_{\nu 23} &={{m}_{12}\over\sqrt{{m}_{12}^2+{m}_{13}^2}}, \\
\label{e:26bb}
s_{\nu 23} &=-{{m}_{13}\over\sqrt{{m}_{12}^2+{m}_{13}^2}},\\
\label{e:26bc}
c_{\nu 12} &=-{{m}_{-}-m_{22}\over 
\sqrt{({m}_{-}-m_{22})^2+{m}_{12}^2+{m}_{13}^2}}
={\sqrt{{m}_{12}^2+{m}_{13}^2}\over 
\sqrt{({m}_+ -m_{22})^2+{m}_{12}^2+{m}_{13}^2}},\\
\label{e:26bd}
s_{\nu 12} &={\sqrt{{m}_{12}^2+{m}_{13}^2}\over 
\sqrt{({m}_{-}-m_{22})^2+{m}_{12}^2+{m}_{13}^2}}
={{m}_{+}-m_{22}\over 
\sqrt{({m}_+-m_{22})^2+{m}_{12}^2+{m}_{13}^2}},
\end{align}
\eseq
where we have used an identity $(m_+-m_{22})(m_--m_{22})=-(m_{12}^2+m_{13}^2)$
for the second equality in the latter two equations.
But, if $m_1=m_+$, we get the opposite for $\theta_{\nu 12}$ as
\bseq
\begin{align}
\label{e:26bax}
c_{\nu 23} &=-{{m}_{12}\over\sqrt{{m}_{12}^2+{m}_{13}^2}}, \\
\label{e:26bbx}
s_{\nu 23} &={{m}_{13}\over\sqrt{{m}_{12}^2+{m}_{13}^2}},\\
\label{e:26bcx}
s_{\nu 12} &=-{{m}_{-}-m_{22}\over 
\sqrt{({m}_{-}-m_{22})^2+{m}_{12}^2+{m}_{13}^2}}
={\sqrt{{m}_{12}^2+{m}_{13}^2}\over 
\sqrt{({m}_+ -m_{22})^2+{m}_{12}^2+{m}_{13}^2}},\\
\label{e:26bdx}
c_{\nu 12} &={\sqrt{{m}_{12}^2+{m}_{13}^2}\over 
\sqrt{({m}_{-}-m_{22})^2+{m}_{12}^2+{m}_{13}^2}}
={{m}_{+}-m_{22}\over 
\sqrt{({m}_+-m_{22})^2+{m}_{12}^2+{m}_{13}^2}}.
\end{align}
\eseq
Then mass eigenvalues, if we demand $m_2^2 > m_1^2$, 
are given by
\bseq
\begin{numcases}
{{\mathbf{m}}_{\rm Dd} =}
{\rm diag}(m_1={m}_{-}, m_2={m}_{+}, m_3={m}_{22}),\ 
\mbox{for }m_{11}+m_{22}>0, \label{e:26bf}\\
{\rm diag}(m_1={m}_{+}, m_2={m}_{-}, m_3={m}_{22}),\ 
\mbox{for }m_{11}+m_{22}<0. \label{e:26bg}
\end{numcases}
\eseq
Note that for $s_{\nu 23}>0$ and $c_{\nu 23}>0$, we should choose the sign
of ${m}_{12}$ and ${m}_{13}$ accordingly. 
In either case, ${m}_{12}$ and ${m}_{13}$ should have opposite signs
such that ${m}_{12}{m}_{13} <0$.

This possibility of having two different choices is the main reason behind
the two different identities between the symmetry parameters and 
$t_{\nu 12}^2$ as in eqs.(\ref{e:43ba})(\ref{e:43bb})
and subsequent equations involving $t_{\nu 12}^2$.
This is also eventually related to the fact that there is no constraint
currently to tell whether $\theta_{23}$ of the PMNS matrix is smaller or
larger than $\pi/4$, even if we restrict the other angle to satisfy
 $\theta_{12}<\pi/4$.

\noindent
\underline{The Physical Dirac Case}

If the observed neutrinos are Dirac types, only
two symmetry constraints eqs.(\ref{e:7a})(\ref{e:7b}) are 
sufficient. As we have seen, the symmetry constrains not only the form 
of the mass matrix but also its components. 
In the pure cyclic case of $C_2(\T')\times C_2(\T)\times C_2(\S)$,
eq.(\ref{e:39}) implies
\beq
\label{e:53}
m_{12}  ={s_\nu c_\alpha\over 1-c_\nu}m_{22},
\eeq
and that, with eq.(\ref{e:40}), we get
\beq
\label{e:54}
m_{13}={s_\nu (1+s_\alpha)\over 1-c_\nu}m_{22},
\eeq
while, with eq.(\ref{e:41}), we get
\beq
\label{e:55}
m_{11}={2c_\nu +s_\alpha(1+c_\nu)\over 1-c_\nu}m_{22}.
\eeq
Then, from eq.(\ref{e:26bf}), we can express the mass eigenvalues
in terms of the symmetry parameters as
\beq
\label{e:m1}
m_1=-m_{22},\quad
m_2={2+s_\alpha(1+c_\nu)\over 1-c_\nu} m_{22},\quad
m_3=m_{22},
\eeq
where $m_{22}$ provides over-all mass scale. 
Under the enhanced symmetry of $C_2(\T')\times D_3(\T,\S)$, there is an
additional constraint, eq.(\ref{e:48x}), which relates $s_\alpha$ to $c_\nu$,
then the masses are more strictly constrained as
\beq
\label{e:m2}
m_1=-m_{22},\quad
m_2={2+c_\nu\over 1-c_\nu}\, m_{22},\quad
m_3=m_{22}.
\eeq
Note that in either cases, $m_1$ and $m_3$ are degenerate, 
which contradicts with the measurement of $\Delta m_{31}^2\neq 0$. 
This situation does not change even with eq.(\ref{e:26bg})
because it only swaps the values of $m_1$ and $m_2$.

Therefore, the symmetry we have imposed disfavors
the physical Dirac neutrinos unless there is a mechanism to break this 
global symmetry to lift the degeneracy. 
However, in the following Majorana case, this degeneracy can be lifted.

\noindent
\underline{The Majorana Case}

In the Majorana case, the physical left-handed neutrino masses are 
due to the see-saw mechanism w.r.t. the right-handed Majorana neutrino masses.
The right-handed Majorana neutrino mass matrix must satisfy eq.(\ref{e:7c}), 
which implies that
\beq
\label{e:34w1}
[\mathbf{m}_R, \S]=0.
\eeq
Since $\S$ is a real symmetric (or hermitian) matrix, 
it can be easily diagonalized as
\beq
\label{e:34w2}
\S_{\rm d}= V^{-1} \S V
\eeq
by an orthogonal (or unitary) matrix $V$.
Then we can easily generate $\mathbf{m}_R$ with three different 
eigenvalues, such that 
$\mathbf{m}_{\rm Rd}={\rm diag}(m_{R1}, m_{R2}, m_{R3})$,
using the similarity transformation that diagonalizes 
symmetric $\S$, as
\beq
\label{e:34w3}
\mathbf{m}_R= V \mathbf{m}_{\rm Rd} V^{-1}
\eeq
such that eq.(\ref{e:34w1}) is satisfied by construction.
So, after the see-saw mechanism, 
we can have non-degenerate left-handed Majorana neutrinos masses given by
\beq
\label{e:34w4}
m_{L1}={m_1^2\over m_{R1}},\quad
m_{L2}={m_2^2\over m_{R2}},\quad
m_{L3}={m_3^2\over m_{R3}},
\eeq
where $m_i$ are from either eq.(\ref{e:m1}) or eq.(\ref{e:m2})
depending on the symmetry. 

In most of cases we consider in this paper, the Dirac neutrino mass eigenvalues
satisfy the ratio $m_1^2 :m_2^2 :m_3^2 \sim 1:3^2:1$, hence the left-handed Majorana mass eigenvalues satisfy 
\beq
m_{L1}^2: m_{L2}^2: m_{L3}^2
\sim {1\over m_{R1}^2}: {81\over m_{R2}^2}: {1\over m_{R3}^2}.
\eeq
For the normal hierarchy, i.e. $m_{L3}^2 \gg m_{L2}^2 \gsim m_{L1}^2$,
the ratio of the right-handed neutrino masses are 
\beq
m_{L1}^2: m_{L2}^2: m_{L3}^2\sim 1:2:32
\sim {1\over m_{R1}^2}: {81\over m_{R2}^2}: {1\over m_{R3}^2},
\eeq
or
\beq
m_{R1}: m_{R2}: m_{R3}\sim 5.7: 36:1\sim 6:36:1,
\eeq
to meet the observed mass relations, eq.(\ref{e:s4mbf}). 
For the inverted hierarchy, i.e. $m_{L2}^2 \gsim m_{L1}^2\gg m_{L3}^2$, 
they are now
\beq
m_{L1}^2: m_{L2}^2: m_{L3}^2\sim 31:32:1
\sim {1\over m_{R1}^2}: {81\over m_{R2}^2}: {1\over m_{R3}^2}
\eeq
or
\beq
m_{R1}: m_{R2}: m_{R3}\sim 1: 8.9:5.6\sim 1:9:6
\eeq
So the symmetry can constrain the relative ratios of 
the right-handed Majorana mass eigenvalues. From the naturalness point of
view, the inverted hierarchy may be slightly favored, but the difference
is not significant.

\newsection{The PMNS Matrix}

Now we have all ingredients and ready to compute the PMNS matrix. 
For this purpose, let us recall and summarize the relations between 
the mixing angles and the symmetry parameters. 
For $C_2(\T')$, $\theta_{\ell 13}$ is given by eq.(\ref{e:10x})
with $M_{11}=M_\mu$, i.e. the ratio in terms of of physical charged 
lepton masses as
\beq
s_{\ell 13}^2 ={M_\mu-M_e\over M_\tau -M_e}\simeq {M_\mu\over M_\tau}.
\eeq
The neutrino part of the mixing angles are such that, 
under pure cyclic groups $C_2(\T)\times C_2(\S)$, from eqs.(\ref{e:43a}),
\beq
s_{\nu 23}^2 ={1+s_\alpha \over 2},
\eeq
and from eqs.(\ref{e:43ba})(\ref{e:43bb}) we have
\bseq
\begin{numcases}
{t_{\nu 12}^2 =} 
{1-c_\nu\over 1+c_\nu}{2\over 1+s_\alpha},\quad \mbox{for } m_1=m_-,\\
{1+c_\nu\over 1-c_\nu}{1+s_\alpha\over 2},\quad \mbox{for } m_1=m_+.
\end{numcases}
\eseq
Under the enhanced symmetry with the dihedral group $D_3(\T,\S)$, 
$s_\alpha$ is no longer independent from $c_\nu$
such that, with eq.(\ref{e:50a}), we have
\beq
s_{\nu 23}^2 = {2c_\nu + 1\over 2(1+c_\nu)}, 
\eeq
and, from eqs.(\ref{e:50b})(\ref{e:50c})
\bseq
\begin{numcases}
{t_{\nu 12}^2 =}{2(1-c_\nu)\over 2c_\nu + 1}, 
\quad \mbox{for } m_1=m_-, \label{e:tnu}\\[6pt]
{2c_\nu+1\over 2(1-c_\nu)},\quad \mbox{for } m_1=m_+. \label{e:tnuinv}
\end{numcases}
\eseq
So the mixing angles are entirely given in terms of the symmetry parameters
and, as the symmetry gets enhanced, they becomes more restrictive.

If we further choose eq.(\ref{e:8y}) such that symmetry parameters of
$\T$ and $\T'$ are related as
\beq
c_\nu=t_{\ell 13},
\eeq
even the symmetry parameters are no longer free at all in the case of 
$C_2(\T')\times D_3(\T,\S)$. So it is much desirable to look for an enhanced
symmetry in which $C_2(\T')$ and $D_3(\T,\S)$ are combined into a single 
group satisfying $c_\nu=t_{\ell 13}$.

The PMNS matrix in our model is given in eq.(\ref{e:1a}) as
\beq
\label{e:25}
U_{\rm PMNS}= U_\ell^\dagger(\theta_{\ell 13}) 
U_\nu(\theta_{\nu 23}, \theta_{\nu 12}),
\eeq
which eventually will only depend on the symmetry parameters.
Using eq.(\ref{e:10}) and eq.(\ref{e:26b}), we can compare the r.h.s.
to the PMNS matrix in the standard (PDG) parametrization as
\beq
\begin{aligned}
\label{e:unu}
U_{\rm PMNS} &=
\left(
\begin{array}{ccc}
c_{12}c_{13} & s_{12}c_{13} &s_{13}\\ 
-(s_{12}c_{23}+c_{12}s_{23}s_{13}) &c_{12}c_{23}-s_{12}s_{23}s_{13} 
&s_{23}c_{13}\\
s_{12}s_{23}-c_{12}c_{23}s_{13} &-(c_{12}s_{23}+s_{12}c_{23}s_{13}) 
&c_{23}c_{13}
\end{array}
\right) \\
&=
\left(
\begin{array}{ccc}
{c}_{\ell 13}c_{\nu 12}+ {s}_{\ell 13}s_{\nu 23}s_{\nu 12}
& {c}_{\ell 13}s_{\nu 12}-{s}_{\ell 13}s_{\nu 23}c_{\nu 12} 
&{s}_{\ell 13}c_{\nu 23}\\ 
-c_{\nu 23}s_{\nu 12} &c_{\nu 23}c_{\nu 12} &s_{\nu 23}\\
-{s}_{\ell 13}c_{\nu 12}+ {c}_{\ell 13}s_{\nu 23}s_{\nu 12}
& -({s}_{\ell 13}s_{\nu 12}+{c}_{\ell 13}s_{\nu 23}c_{\nu 12})
&{c}_{\ell 13}c_{\nu 23}
\end{array}
\right).
\end{aligned}
\eeq
Then we can obtain the mixing angles of the PMNS matrix in terms of the
our theoretical mixing angles of charged lepton and Dirac neutrino mass 
matrices as
\bseq
\begin{align}
\label{e:28a}
s_{13} &={s}_{\ell 13} c_{\nu 23},\\
t_{23} &={t_{\nu 23}\over {c}_{\ell 13}},\\
t_{12} &={t_{\nu 12} -{t}_{\ell 13} s_{\nu 23} \over
1+{t}_{\ell 13} s_{\nu 23}t_{\nu 12}}.
\end{align}
\eseq
Eq.(\ref{e:28a}) implies that the smallness of $s_{13}$ is related to 
the smallness of the muon-to-tau mass ratio. 

There are three symmetry parameters: $\theta_\ell=\theta_{\ell 13}/2$ for 
$C_2(\T')$,
$\theta_\nu$ for $C_2(\T)$, and additional $\theta_\alpha$ for $C_2(\S)$.
Although our main interest is the most restrictive case (Case IV below), 
but for comparison purpose we will consider the following four different cases: 

$\bullet$ Case I: $C_2\times C_2\times C_2$ with all independent 
$\theta_\ell $, $\theta_\nu$ and $\theta_\alpha$.

$\bullet$ Case II: $C_2\times C_2\times C_2$ with $c_\nu=t_{\ell 13}$, but
independent $\theta_\alpha$.

$\bullet$ Case III: $C_2\times D_3$ with only $\theta_\ell $ and $\theta_\nu$
independent.

$\bullet$ Case IV: $C_2\times D_3$ with $c_\nu=t_{\ell 13}$ such that
all symmetry parameters are related.

Note that, in the following, unless specifically mentioned, 
we use eq.(\ref{e:tnu}) for $t_{\nu 12}$.

\noindent
\underline{Case I}

With simple choices
$s_{\ell 13}^2=\tilde{M}_\mu/\tilde{M}_\tau\simeq M_\mu/M_\tau\simeq 0.05919$,
$c_\nu=1/4$ and $s_\alpha=1/5$ such that $s_{\ell 23}^2=3/5$ and 
$t_{\nu 12}^2=1$, we can obtain 
$s_{12}^2\simeq 0.313$, $s_{13}^2\simeq 0.0237$ and
$s_{23}^2\simeq 0.615$. These agree with the best-fit values given in
eq.(\ref{e:s4bf}) impressively well, albeit with $\theta_{23}>\pi/4$. 

Note that the current observed data do not provide
any information whether $\theta_{23}$ is smaller or larger than $\pi/4$,
so $s_{23}^2\simeq 0.61$ is equally acceptable as $s_{23}^2\simeq 0.39$.
Nevertheless, we can obtain $s_{23}^2\simeq 0.4$, too. For example, 
if we choose $c_\nu\simeq 0.55$, $s_\alpha\simeq -0.24$ and
$s_{\ell 13}^2\simeq 0.04$ such that $t_{\nu 12}^2\simeq 0.76$ and 
$s_{\nu 23}^2\simeq 0.38$, then we obtain $s_{23}^2\simeq 0.39$, 
$s_{12}^2\simeq 0.312$ and $s_{13}^2\simeq 0.0248$. However, 
$s_{\ell 13}^2$ needs to be much smaller to obtain the needed $s_{13}^2$,
and no longer related to the muon-to-tau mass ratio.

If we assume $c_\nu=1/3$ and $s_\alpha=0$ to get $s_{\nu 23}^2=1/2$ and
$t_{\nu 12}^2=1$, i.e. BM (bimaximal)\cite{Barger:1998ta}.
For $s_{\ell 13}^2\simeq M_\mu/M_\tau\simeq 0.05919$, we obtain
$s_{12}^2\simeq 0.328$, $s_{13}^2\simeq 0.0296$ and
$s_{23}^2\simeq 0.515$, but $s_{13}^2$ about $2\sigma$ away and $s_{23}^2$
is more than $\sigma$ away. Even if we relax $s_{\ell 13}^2$ to different
values, we cannot get better results. 
So, although it may not be ruled out, it is not as good as other cases 
we consider.

\noindent
\underline{Case II}

In this case, $c_\nu$ is no longer independent but given by eq.(\ref{e:8y}),
$c_\nu =t_{\ell 13}$. Then, 
for $s_{\ell 13}^2\simeq M_\mu/M_\tau\simeq 0.05919$, 
we have $c_\nu \simeq 0.2508$.
Now we choose $s_\alpha=1/5$ such that $s_{\nu 23}^2\simeq 3/5$ 
and $t_{\nu 12}^2=0.998$. Then we obtain
$s^2_{12}\simeq 0.312$, $s^2_{23}\simeq 0.615$ and $s^2_{13}\simeq 0.0237$.

If we relax $M_{11}$ so that $M_{11}\neq M_\mu$ is allowed, then we can 
choose $c_\nu=1/4=t_{\ell 13}$, which leads to $s_{\ell 13}=0.2425$
Then, with $s_\alpha=1/5$, we can obtain $s_{\nu 23}^2=3/5$, and that
$s^2_{12}\simeq 0.313$, $s^2_{23}\simeq 0.614$ and $s^2_{13}\simeq 0.0235$.

Note that, compared to the $s_{23}^2\simeq 0.6$ case, a larger $c_\nu$ is 
needed to obtain $s_{12}^2\simeq 0.31$ if $s_{\nu 23}^2$ and $s_{23}^2$ 
are smaller. But, $c_\nu=t_{\ell 13}$ forbids larger $c_\nu$ without
making $s_{13}^2$ become larger, 
so Case II favors $s_{23}^2\simeq 0.6$.

\noindent
\underline{Case III}

Once the symmetry $C_2(\T)\times C_2(\S)$ is enhanced to $D_3(\T,\S)$,
$s_\alpha$ is no longer independent so that $s_{\ell 13}$ and $c_\nu$ 
can control all mixing angles.

For $s_{\ell 13}^2\simeq M_\mu/M_\tau\simeq 0.05919$, if we choose
$c_\nu=1/4$, then $s_{\nu 23}^2= 3/5$ and $t_{\nu 12}^2=1$
so that we can obtain 
$s^2_{12}\simeq 0.313$, $s^2_{23}\simeq 0.615$ and $s^2_{13}\simeq 0.0237$.

One interesting case which may have more geometrical connection is to choose 
$\theta_\nu = 75^\circ$ such that
$c_\nu=(\sqrt{3}-1)/2\sqrt{2}\simeq 0.2588$, then
$s_{\nu 23}^2\simeq 0.603$ and $t_{\nu 12}^2\simeq 0.977$
so that we can obtain 
$s^2_{12}\simeq 0.307$, $s^2_{23}\simeq 0.617$ and $s^2_{13}\simeq 0.0235$.
$75^\circ$ happens to be an half of the inner angle of a dodecagon,
but at this moment we do not know if this has any significance from the
symmetry point of view. 

\noindent
\underline{Case IV}

Now even $c_\nu$ is no longer independent but given by 
$c_\nu =t_{\ell 13} \simeq 0.2508$
for $s_{\ell 13}^2\simeq M_\mu/M_\tau\simeq 0.05919$.
Then $s_{\nu 23}^2\simeq 0.6$ and $t_{\nu 12}^2\simeq 0.998$ such that
we obtain
$s^2_{12}\simeq 0.312$, $s^2_{23}\simeq 0.615$ and $s^2_{13}\simeq 0.0237$.
Physically, this is the most plausible case because $\tilde{\theta}_{13}$
is related to the muon-to-tau mass ratio. However, it is hard to see
if the symmetry we impose has any connection to a geometry, which often
enables us to justify the origin of the symmetry. 

For a close simple fraction, if we choose $c_\nu=1/4$,
which leads to $s_{\ell 13}=0.2425$, $s_{\nu 23}^2=3/5$
and $t_{\nu 12}^2=1$, then we obtain
$s^2_{12}\simeq 0.313$, $s^2_{23}\simeq 0.614$ and $s^2_{13}\simeq 0.0235$.

If we choose $\theta_\nu = 75^\circ$, then 
$c_\nu=(\sqrt{3}-1)/2\sqrt{2}\simeq 0.2588$ and $s_{\ell 13}\simeq 0.2506$. 
We get $s_{\nu 23}^2\simeq 0.603$ and $t_{\nu 12}^2 \simeq 0.977$. 
This leads to
$s^2_{12}\simeq 0.301$, $s^2_{23}\simeq 0.618$ and $s^2_{13}\simeq 0.0249$.

So far, all the examples in Case IV favor $s_{23}^2\simeq 0.6$, and we have
already noticed in Case I that we have to make quite difference choices to 
obtain $s_{23}^2\simeq 0.4$. In Case IV we can also find out
what kind of choice we have to make to obtain $\theta_{23}<\pi/4$ 
and at what expense. From eq.(\ref{e:tnuinv})
if we let $s_{\ell 13}=-\sqrt{0.036}\simeq -0.190$, i.e. 
$c_\nu= t_{\ell 13}\simeq -0.193$, then
$s_{\nu 23}^2\simeq 0.38$ and $t_{\nu 12}^2\simeq 0.257$ so that we can obtain
$s^2_{23}\simeq 0.389$, $s^2_{13}\simeq 0.022$ and $s^2_{12}\simeq 0.308$.
Again, $\theta_{\ell 13}$ is no longer related to the muon-to-tau 
mass ratio.

Some of these are summarized in the Table \ref{t:1} for a quick view.

\begin{table}[tdp]
\begin{center}
\begin{tabular}{|c|c|c|c|c|c|c|}
\hline
Case& $s_{\ell 13}^2$&$c_\nu$ & $s_\alpha$ 
& $\sin^2(2\theta_{12})$ &$\sin^2(2\theta_{23})$ & $\sin^2(2\theta_{13})$ \\
\hline
&&& &$0.857{+0.023\atop -0.025}$ & $>0.95$&  $0.095\pm 0.010$  \\ \hline
I& $M_\mu/M_\tau$ &1/4 &1/5 &0.860 &0.948 &0.0925 \\ \hline
II&$M_\mu/M_\tau$&(0.2508) &1/5 &0.859 & 0.948&  0.0925  \\ \hline
III&$M_\mu/M_\tau$  &1/4 &(1/5) &0.860 &0.948 &0.0925 \\ \hline
IV&$M_\mu/M_\tau$  &(0.2508)  &(0.201)&0.859 &0.947 &0.0924 \\ \hline
IV&(0.0588) &1/4 &(1/5) &0.861 & 0.948&  0.0919  \\ \hline
\end{tabular}
\end{center}
\caption{Comparison between our prediction and the best fit given by 
\cite{pdg2013}. For $C_2\times C_2\times C_2$, Case I is with independent 
$c_\nu$ and $s_\alpha$, Case II is with $c_\nu=t_{\ell 13}$. 
For $C_2\times D_3$, Case III
is with independent $c_\nu$, Case IV is with $c_\nu=t_{\ell 13}$.
The numbers in brackets for $s_{\ell 13}^2$, $c_\nu$ and $s_\alpha$,
are dependent numbers.}
\label{t:1}
\end{table}%

\newsection{Discussions}

We have shown that there is a new discrete symmetry which leads closely to 
the observed PMNS matrix with the small $s_{13}^2$ being related to the 
small muon-to-tau mass ratio (to be precise, if $\theta_{23}>\pi/4$). 
In the absence of the CP-violation, the combined number of parameters in both
(hermitian) mass matrices are twelve, and all of them are fixed by, at most, three symmetry parameters, depending on the symmetry. In the most
restrictive case of $C_2(\T')\times D_3(\T, \S)$ with $c_\nu=t_{\ell 13}$,
just one symmetry parameter $\theta_\ell=2\theta_{\ell 13}$ ends up controlling
all twelve parameters of mass matrices and three mixing angles of 
the PMNS matrix. In other words, one charged
lepton mixing angle $\theta_{\ell 13}$, which itself is given in terms of the
physical charged lepton masses as 
$\theta_{\ell 13}=\sin^{-1}\sqrt{(M_\mu-M_e)/(M_\tau-M_e)}
\simeq \sin^{-1}\sqrt{M_\mu/M_\tau}$\cite{La:2013gga}, 
actually controls all parameters. So there is no free parameter and 
no fine-tuning involved, hence it should be sufficient for us to call 
our production of mixing angles as a prediction, which happens to match
the currently observed data extremely well.

In this paper, the choice $c_\nu=t_{\ell 13}$ 
is not based on any symmetry relations, so it will be very interesting 
to find out if there is a larger symmetry enhanced from 
$C_2(\T')\times D_3(\T, \S)$, under which this shows up as an identity.
We believe such a symmetry exists.

Based on the success of the symmetry we have used in this paper, 
understanding the origin of this $C_2(\T')\times D_3(\T, \S)$ 
from more fundamental point of view could be the key to find the correct 
model to go beyond the SM. 
Since our symmetry generating group elements have negative
determinants, our discrete group cannot be a subgroup of compact gauge
symmetry groups. However, this type of finite groups might rise from a
lattice compactification of string theory. So it will be interesting to
investigate such a possibility.

There is an interesting aspect of our model on charged lepton masses
because our model also relates charged lepton masses to the symmetry.
The charged lepton mass matrix given in eq.(\ref{e:2a}) with
$M_{11}=M_\mu$ has largest mass ratio only about 16, but it reproduces
the large mass hierarchy $M_\tau/M_e\sim 10^3$. So, in some sense, the
lepton mixing under $C_2(\T')$ justifies the lepton mass hierarchy. 
Furthermore, if the tau mass had not been known, we could have actually 
predicted it based on the symmetry we impose here. For example, 
let $c_\nu=t_{\ell 13}=1/4$ and $M_{11}=M_\mu$,
then we can estimate $M_\tau$ in terms of $M_e$ and $M_\mu$ as
$M_\tau=17M_\mu-16M_e\simeq 1788$ MeV. This is less than 1 $\%$ off
from the measured $M_\tau\simeq 1776.82\pm 0.16$ MeV. 
Even though it is still $70\sigma$ away, the outcome is quite intriguing. 
It will be interesting to find out if there is any deeper meaning to this.
For example, $\theta_{\ell 13}=\tan^{-1}(1/4)$ could be 
satisfied with the tree level values of charged leptons, 
or at some energy scale.

Another interesting aspect of our model is that the symmetries we impose
favor Majorana neutrinos because they symmetries force Dirac neutrino masses
be degenerate. As is well known, the Majorana neutrinos are also favored 
because we do not have to assume unnaturally small Yukawa coupling constants 
unlike the Dirac case. The symmetry we impose also puts a constraint on
the right-handed neutrino mass matrix such that $[\S, \mathbf{m}_R]=0$.

In this paper, we have not considered the CP-violation, but it should
be straightforward to include it. 
For example, in the case for $s_{\ell 13}=-\sqrt{0.036}<0$ to obtain 
$\theta_{23}<\pi/4$, we actually get $s_{13}<0$ so that  
this can be interpreted as $s_{13}>0$ with $\delta=\pi$.
Furthermore, since our model favors the Majorana neutrinos,
there should be three CP-violating phases, which can be easily introduced
because our symmetry generating transformation matrices can be all hermitian.
We will leave details of this as a future work.

\noindent
{\bf Acknowledgments:}
I would like to thank Tom Weiler for numerous conversations on related subjects,
Tom Kephart for conversations on the discrete groups,
and David Ernst for his interests in this approach.

\renewcommand{\Large}{\large}

\end{document}